\documentstyle[12pt,epsfig]{article}

\newcommand{\e}{\epsilon}
\newcommand{\coe}{{\bar{\epsilon}}}

\newcommand{\th}{\mbox{tanh}}

\newcommand{\f}{\begin{equation}}
\newcommand{\ff}{\end{equation}}
\newcommand{\beq}{\begin{equation}}
\newcommand{\eeq}{\end{equation}}
\newcommand\bea{\begin{eqnarray}}
\newcommand\eea{\end{eqnarray}}

\newcommand{\id}{\mbox{id}}

\makeatletter
\newcommand{\ps@preprint}{%
        \renewcommand{\@oddhead}{\hfil\small{CGPG-00/6-4, 
        Imperial/IC/99-00/30}}}%
\makeatother
\begin{document}
\title{An algebraic approach to coarse graining}
\author{Fotini Markopoulou\thanks{Email address: fotini@ic.ac.uk.}\\
	    Center for Gravitational Physics and Geometry, \\
	    Department of Physics,\\
	    The Pennsylvania State University,\\
	    University Park, PA 16801, USA.\\
	    and\\
	    The Blackett Laboratory, \\
            Imperial College of Science, Technology and Medicine,\\
            Prince Consort Road, South Kensington, \\
	    London SW7 2BZ, U.K.}
\date{June 25, 2000}
\maketitle
\thispagestyle{preprint}  
\vfill
\begin{abstract}
    We propose that Kreimer's method of Feynman diagram
    renormalization via a Hopf algebra of rooted trees 
    can be fruitfully employed in the analysis of block spin
    renormalization or coarse graining of inhomogeneous
    statistical systems.  Examples of such systems include
    spin foam formulations of non-perturbative quantum gravity
    as well as lattice gauge and spin systems on irregular
    lattices and/or with spatially varying couplings. 
    We study three examples which are $Z_2$ lattice gauge
    theory on irregular 2-dimensional lattices, Ising/Potts 
    models with varying bond strengths and  
    $(1+1)$-dimensional spin foam models.
\end{abstract}
\vfill
\newpage

\section{Introduction}

Spin foam models are a natural description of non-perturbative quantum 
gravity, both Lorentzian \cite{lorentzianSF,ALA} and euclidean 
\cite{euclideanSF,FeynmanSF}.  In these models, we need to sum over 
discrete spacetime histories and, on each history, perform a sum over 
labels which live in the space of representations of some compact or 
quantum group (usually $SU(2)$ or $SU_q(2)$).  Thus, for each history, 
the sum to be performed defines a generalization of a spin system or a 
lattice gauge theory.  At the same time, the sum over all histories 
may be thought of as a sum over Feynman diagrams, in which case the 
sums over labels replace the momentum integrals (see, for example, 
\cite{FeynmanSF}).

The main problem in spin foam models is finding those theories which 
have a good continuum limit, in the sense that coarse-grained 
observables, averaged over many Planck lengths, describe a classical 
spacetime.   While the combinatorial part of the problem of spin 
foam renormalization is similar to that of Feynman diagrams, the 
physical limits differ: in spin foam renormalization there is a 
physical cutoff, the Planck scale, and the interesting limit is that 
of a very large number of vertices, corresponding to an spacetime 
volume which is large in Planck units.  Thus, the continuum limit of 
spin foam models should be formulated as a renormalization group 
problem.

The main motivation of the present work is to develop such a 
renormalization group approach for spin foam models.  A particular 
spin foam can be thought of as a generalization of a spin system or a 
lattice gauge system, however, an inhomogeneous one, as it is an 
irregular 2-complex with varying bond strengths.  Even if the bond 
strengths are taken to be the same initially, they will vary after a 
few block transformations of the irregular complex.  A central 
ingredient in the solution of the renormalization problem for spin 
foams will then have to be a method to renormalize inhomogeneous 
systems.  These are systems defined on irregular lattices, systems in 
which the bond strengths vary in space, or systems with both 
characteristics.  It is more difficult to apply renormalization group 
techniques to such systems because there may be no bond strength which 
characterizes a particular scale.  In this note we suggest that, for 
such systems, a generalization of the renormalization group exists, 
which is based on a Hopf algebra.  The starting point is the recent 
discovery by Kreimer of a Hopf algebra which underlies quantum field 
theory renormalization \cite{Kreimer}, and has led to a new approach 
to perturbative renormalization which is currently under 
development \cite{KHA}.

We show here that a similar method applies to inhomogeneous spin 
systems and spin foams.  As examples, we study the renormalization of 
the $Z_2$ lattice gauge theory on an inhomogeneous 2-dimensional 
lattice, and the 1-dimensional Ising model (2-dimensional Ising/Potts 
models are a straightforward generalization).  We then give the 
renormalization operations on the partition function of a generic 1+1 
spin foam.

The basic idea is that the (partitioned) spin systems are the elements 
of the algebra, and block transformations, which change a spin system 
to another by integrating over subsets of spins, are operations on 
these elements.  Among these elements are homogeneous systems.  The 
standard renormalization group transformation changes a homogeneous 
system to another homogeneous system, and corresponds to the 
homogeneous coupling terms in a particular operation of the Hopf 
algebra.  Under this new light, it is intriguing to consider the 
statement that the renormalization group is not a group because there 
is no inverse.  A Hopf algebra is also not a group, but there is an 
operation which behaves as a generalized inverse, called the antipode.  
What we find here is that the renormalization group transformation of 
a spin system is closely related to the antipode.  This is a 
non-perturbative analogue of Kreimer's results for renormalizable 
field theory, in which the antipode of a diagram is related to the 
counterterms necessary to subtract its divergencies.

In the next section, we define the Hopf algebra associated with the 
block transformations of the $Z_2$ lattice gauge theory on a 
2-dimensional lattice.  The renormalization of the theory via this 
algebra is illustrated for a simple example.  In section 3, we give a 
simpler example of Hopf algebra block transformations, the 
1-dimensional Ising model.  Then, in section 4, we treat the formal 
partition function of spin foams in 1+1 dimensions.  We define the Hopf 
algebra of partitioned spin foams and the general form of a 
coarse-graining operation.  In the concluding section we indicate ways 
in which these results may be used.  The properties of a Hopf algebra 
can be found in \cite{Hopf}, and for rooted trees in \cite{Kreimer}.


\section{Coarse-graining the $Z_2$ lattice gauge theory on a 
2-dimensional lattice}

In this section we study a simple example of the generalized 
renormalisation group transformation, carried out by the Hopf algebra 
of parenthesized Boltzmann weights.

\begin{figure}
\centerline{\mbox{\epsfig{file=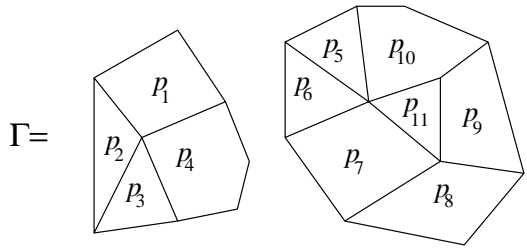}}}
\caption{A 2-dimensional lattice with its plaquettes marked.} 
\label{lattice}
\end{figure}

We start by giving a 2-dimensional lattice $\Gamma$ (which may be 
connected, or a set of disjoint lattices), with its edges labeled by 
$Z_2$ elements $q_i=\pm 1$.  We call {\em plaquettes} the smallest 
loops in the lattice, namely, loops that contain no further loops.  
Each plaquette $p$ is labeled by a coupling $\kappa_p$, a real number 
(figure \ref{lattice}).

A particular configuration (assignment of group elements and 
couplings) on the lattice $\Gamma$ gives rise to a Boltzmann weight
\beq
w_\Gamma=\exp\left(\sum_{p\in \Gamma}\kappa_p \prod_{q_i\in p} 
q_i\right),
\label{eq:w}
\eeq
which is a function of the labels on the lattice.   
The product ranges over the labels on  all the edges around the 
plaquette $p$.  The temperature and Boltzmann's constant
have been absorbed in the remaining parameters.   

The partition function for the $Z_2$ lattice gauge theory is  
the sum of the weights over all assignments of edge labels on $\Gamma$:
\beq
Z(\{\kappa_p\})=\sum_{\{q_i\} }
\exp\left(-\sum_{p\in\Gamma} \kappa_p \prod_{q_i\in p} q_i\right).
\eeq

A particular case is when $\Gamma$ is a regular lattice, for example, 
a square one.  Then all the plaquettes are squares with sides 
of length $a$ and all the couplings $\kappa_p$ 
are the same.  A renormalization group transformation
on this system is a homogeneous coarse-graining of $\Gamma$ to a new
square lattice $\Gamma'$, with lattice spacing, say, 
$a'=2 a$.
Such a transformation may be performed by following these steps:
First, partition $\Gamma$ into square sublattices $\gamma$, each with 
side $2a$.  In each $\gamma$, sum over all labels on the 
internal edges, to obtain a 
new larger square plaquette $p'$ labeled by a coupling $\kappa_{p'}$:
\beq
\kappa_{p'}=2^{4} \tanh^{-1}\left(\tanh ^4\kappa_p\right).
\label{eq:RG1}
\eeq
(For a derivation of this equation, see section 3.)
Finally, we redefine the edges of the new plaquette so that $p'$ is a  
square with new sides $q_1', q_2', q_3', q_4'$, each a product of the 
labels on two old sides. 
It contributes $\exp(-\kappa_{p'} q_1' q_2' q_3' q_4')$ to the partition 
function. 

We repeat this procedure until we reach the desired new lattice 
spacing.  The result is the renormalization group equation, the 
generalization of (\ref{eq:RG1}) to all the plaquettes on the final
lattice.  It has the form  
\beq
\{\kappa'\}=R\left(\{\kappa\}\right),
\label{eq:RG2}
\eeq
which provides the couplings $\{\kappa'\}$ 
on the new lattice $\Gamma'$ in terms of the couplings $\{\kappa\}$
on $\Gamma$.

Of course, summing over all the internal labels is only one possible 
coarse-graining scheme.  Other schemes, 
exact or approximate (such as decimation or truncation), can be 
used as long as they respect the gauge invariance of the theory.  

Our aim in this section is to 
generalise this coarse-graining procedure to inhomogeneous 
lattices in a way that gives rise to a Hopf algebra 
and an extension of the renormalisation group equation 
(\ref{eq:RG2}).

\subsection{The Hopf algebra of partitioned $Z_2$ lattices and their 
weights}

First we need to formalize the correspondence between Boltzmann weights 
and the underlying lattices.  
We call a lattice $\Gamma$ {\em partitioned} when it is marked with an 
{\em allowed}  partition into a set of sublattices.  An allowed 
partition contains no overlapping sublattices.  Namely, for any two 
sublattices $\gamma_1,\gamma_2$ in the partition, either 
$\gamma_1\subset\gamma_2$, $\gamma_2\subset\gamma_1$, or, 
$\gamma_1\cap\gamma_2=\emptyset$.  A sublattice may be a set of 
disconnected non-overlapping sublattices.

Let us call $V$ the space of labeled partitioned 
lattices over the real numbers.  A general element in $V$ is a sum of 
labeled partitioned lattices, with real coefficients. 
 We can turn $V$ into an algebra by defining 
{\em multiplication}, $m:V\otimes V\rightarrow V$,
to be the disjoint union of two lattices:
\beq
\Gamma_1\cdot\Gamma_2=\Gamma_1\cup\Gamma_2.
\eeq
We have written $\Gamma_1\cdot\Gamma_2$ for $m(\Gamma_1\otimes\Gamma_2)$. 
The unit element is the empty lattice $e$. 
The {\em unit} operation, $\epsilon:R\rightarrow V$, turns a real 
number into a lattice by multiplying it 
with the empty lattice, $\epsilon(r)=re,\ r\in R$.
The generators in $V$ are the connected labeled partitioned 
lattices $\{\Gamma_c\}$.

We will now define the rest of the operations needed for $V$ to be a 
Hopf algebra.  First, let $\gamma$ denote a proper
sublattice of $\Gamma$, namely $\gamma\neq e$ and $\gamma\neq 
\Gamma$.   We call the lattice that remains if 
we ``cut out'' $\gamma$ from $\Gamma$, the {\em remainder}, and denote 
it by $\Gamma/\gamma$.  That is, in the lattice
\beq
	\begin{array}{c}\mbox{\epsfig{file=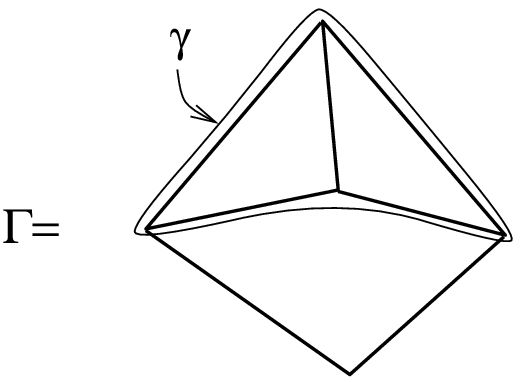}}\end{array}
\eeq
with the marked sublattice $\gamma$, the remainder is\footnote{
This definition of the remainder is the simplest one when the weight 
on a lattice factorizes into weights on its sublattices in the 
partition and the renormalization scheme preserves this.  Otherwise, 
the remainder should be defined to have the same external edges 
as the original.  This is the case for the remainder in Section 4 for 
a spin foam, which is also the same as in \cite{Kreimer,KHA}.
}
\beq
	\begin{array}{c}\mbox{\epsfig{file=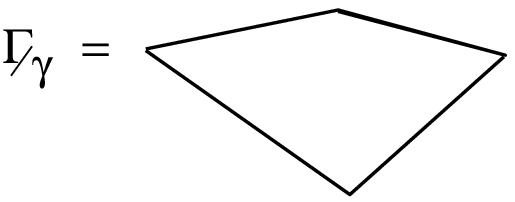}}\end{array}
\eeq

The {\em coproduct} in $V$ is the operation $\Delta:V\rightarrow V\otimes V$
defined by
\begin{eqnarray}
    \Delta(\Gamma) &= & \Gamma\otimes e+ e\otimes\Gamma+
             \sum_\gamma \gamma\otimes{\Gamma}/{\gamma}
	     \label{eq:coproduct}
	     \\
    \Delta(e)&=&e\otimes e   \\
    \Delta(\Gamma_1\cdot\Gamma_2)&= & \Delta(\Gamma_1)\Delta(\Gamma_2).
\end{eqnarray}
The sum in (\ref{eq:coproduct}) ranges over all sublattices $\gamma$ 
{\em in the given partition} of $\Gamma$.  
Note that sublattices $\gamma_p$ with no further sublattices 
are special.  They satisfy 
\beq
\Delta(\gamma_p)=\gamma_p\otimes e+\e\otimes\gamma_p.
\eeq
These are the {\em primitive elements} of the Hopf 
algebra.  Plaquettes are always primitive. 

The {\em counit} is an operation $\coe:V\rightarrow R$ that annihilates 
every lattice except $e$:
\beq
\coe(\Gamma)=
\left\{ \begin{array}{lr}
0& \mbox{for }\Gamma\neq e,\\
1& \mbox{for }\Gamma=e.
\end{array}\right.
\eeq
The counit and the coproduct satisfy 
$(\id\otimes\epsilon)\Delta(\Gamma)= 
(\epsilon\otimes\id)\Delta(\Gamma)= \Gamma$.  
One can check that this coproduct is coassociative, i.e.\ it satisfies 
$(\Delta\otimes\id)\Delta=(\id\otimes\Delta)\Delta$.
It is not cocommutative, namely, if we switch the order of the lattice 
pairs in each term in (\ref{eq:coproduct}) we get an element 
of $V\otimes V$ different from $\Delta(\Gamma)$. 

By adding the $\Delta$ and $\coe$ operations to $V$, we have turned 
it into a bialgebra which is associative, coassociative and 
commutative.  We now only need to define an {\em antipode} for $V$ to be  a 
Hopf algebra.  An antipode is an operation $S:V\rightarrow V$, that 
satisfies 
\beq
m(S\otimes \id)\Delta(\Gamma) = \epsilon\coe(\Gamma)=
\left\{ \begin{array}{lr}
0& \mbox{for }\Gamma\neq e,\\
e& \mbox{for }\Gamma=e.
\end{array}\right.
\label{eq:Sdefinition}
\eeq 
as well as $m(\id\otimes 
S)\Delta(\Gamma)= \epsilon\coe(\Gamma)$.  We will need this later and 
so we name $O$ the LHS of (\ref{eq:Sdefinition}):
\beq
O(\Gamma)=
m(S\otimes \id)\Delta(\Gamma) =\left\{ \begin{array}{lr}
0& \mbox{for }\Gamma\neq e,\\
e& \mbox{for }\Gamma=e.
\end{array}\right.
\label{eq:O}
\eeq 

The antipode on the partitioned lattices is
\begin{eqnarray}
S(\Gamma)&=&-\Gamma-\sum_\gamma S(\gamma){\Gamma}/{\gamma}
\label{eq:S}\\
S(\gamma_p)&=&-\gamma_p\\
S(e)&=&e\\
S(\Gamma_1\cdot\Gamma_2)&=&S(\Gamma_1) S(\Gamma_2).
\end{eqnarray}
$S(\Gamma)$ is an iterative equation that stops when a primitive 
lattice $\gamma_p$ is reached. 

One can check that the operation (\ref{eq:S}) applied to a partitioned 
lattice $\Gamma$ satisfies the (\ref{eq:Sdefinition}).  We thus have 
a Hopf algebra of partitioned labeled lattices.  

That the lattices we use are partitioned is important because the 
partition dictates the order in which we will do the block spin 
transformation.  However, the transformation is really an operation 
on the Boltzmann weight on a lattice.  
It is possible to encode the partitioning of a 
lattice in its Boltzmann weight by using {\em parenthesized} weights.  
This means writing the weight as a product the weights of each 
plaquette in the lattice, $w_\Gamma=\prod_p 
e^{\kappa_p\prod q_i}$, and marking by brackets the factors in the 
product that correspond to the sublattices in the given partition.

The rules are the following.  The weight of a single 
sublattice $\gamma$ is enclosed in a set of matching open/closing 
brackets 
$(w_\gamma)$.  Two nested sublattices, $\gamma_1\subset\gamma_2$ are 
put in nested brackets: $((w_{\gamma_1})w_{\gamma_2})$.  Two disjoint 
sublattices are marked by a disjoint pair of brackets: 
$(w_{\gamma_1})(w_{\gamma_2})$.  A single plaquette lives in a single set 
of brackets, with no nesting inside.  Finally, if $\Gamma$ is a 
connected lattice, the outermost left bracket matches the outermost 
right one\footnote{These rules are different than the parenthesized 
words of Kreimer in \cite{Kreimer}, since the weight of a lattice 
factorizes into weights of its sublattices, and we use an exact 
renormalization scheme which respects this property.}.  

For example, the parenthesized weight of the partitioned lattice 
\beq
	\begin{array}{c}\mbox{\epsfig{file=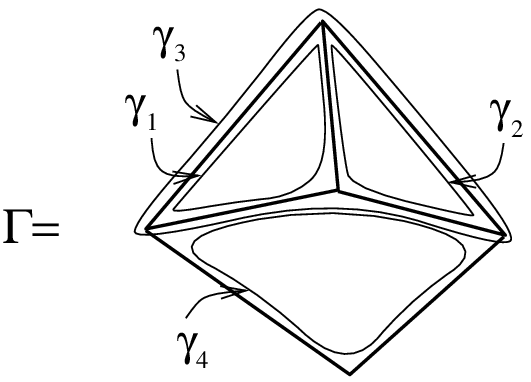}}\end{array}
	    \label{eq:example1}
\eeq
with labels
\beq
	\begin{array}{c}\mbox{\epsfig{file=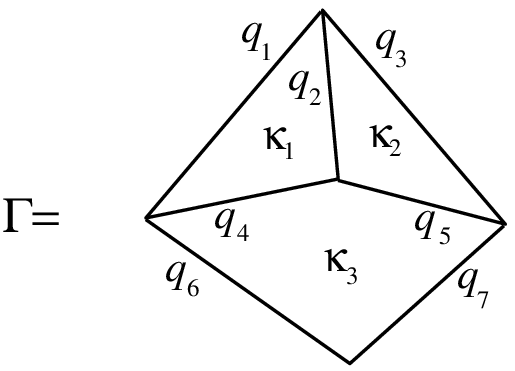}}\end{array}
	    \label{eq:labeled}
\eeq
is
\beq
\begin{array}{rl}
w_\Gamma=&\left(\left(\left(w_{\gamma_1}\right)\left(w_{\gamma_2}\right)
\right)
     \left(w_{\gamma_4}\right)\right)\\
=&
\left(\left(\left( e^{\kappa_{p_1}\prod_{q_i\in p_1} q_i}\right)
\left( e^{\kappa_{p_2}\prod_{q_i\in p_2} q_i}\right)\right)
\left( e^{\kappa_{p_3}\prod_{q_i\in p_3} q_i}\right)\right).
\end{array}
\label{eq:wEX1}
\eeq
From now on, when we write $\Gamma$ we mean a (labeled) partitioned 
lattice, and by $w_\Gamma$ we will mean a parenthesized  weight.  

Therefore, to a partitioned labeled lattice corresponds a 
parenthesized weight, which is a real function of the labels on the 
lattice, of the form (\ref{eq:w}), marked with a bracket structure 
according to the above rules.  In fact, we can think of a weight $w$ 
as a map from $V$ to the algebra of such parenthesized weights, which we 
will call $W$.  All the operations we 
defined for $V$ also apply to weights in $W$.  Let us list them. 

A general element in $W$ is a sum of parenthesized weights with real 
coefficients.  Multiplication in $W$ is the product of the 
weights on two disjoint lattices: 
\beq
w_{\Gamma_1}\cdot w_{\Gamma_2}=w_{\Gamma_2\cdot\Gamma_2}.
\eeq
The unit element in $W$ is 1, which we define to  be the weight of the 
empty lattice: $w_e=1$. 
The unit operation $\epsilon$ takes a real number $r$ to $r \/ w_e$.
Again, there is a set of generating elements, the weights on 
connected lattices. 

The coproduct is 
\bea
\Delta(w_\Gamma)&=&w_\Gamma\otimes 1+1\otimes w_\Gamma
+\sum_\gamma w_{\gamma}\otimes w_{\Gamma/\gamma}\\
\Delta(1)&=&1\otimes 1\\
\Delta(w_{\Gamma_1} w_{\Gamma_2}) &=& 
\Delta(w_{\Gamma_1})\cdot \Delta(w_{\Gamma_2}). 
\eea
$w_{\Gamma/\gamma}$ is the weight of the remainder, equal to 
$\frac{w_\Gamma}{w_\gamma}$.  

Finally, the antipode is given by 
\bea
S(w_\Gamma)&=&-w_\Gamma-\sum_\gamma S(w_\gamma) w_{\Gamma/\gamma}\\
S(w_{\gamma_p})&=&-w_{\gamma_p}\\
S(1)&=&1\\
S(w_{\Gamma_1}w_{\Gamma_2})&=&S(w_{\Gamma_1}) S(w_{\Gamma_2}).
\eea
It satisfies $O$ given in (\ref{eq:O}).

\subsection{Example}

Let us now give an example of the coproduct and antipode operations on 
a partitioned lattice.  

Consider the lattice (\ref{eq:labeled}), with the partition
(\ref{eq:example1}).
The coproduct on this lattice  produces all possible pairs of sublattices and 
remainders in the given partition.  It is 
\beq
\begin{array}{rl}
    \Delta(\Gamma)=& \Gamma\otimes e+e\otimes\Gamma+
    \gamma_1\otimes\Gamma/\gamma_1+ \gamma_2\otimes\Gamma/\gamma_2\\
    &+\gamma_3\otimes\Gamma/\gamma_3+ \gamma_4\otimes\Gamma/\gamma_4\\
    =&\Gamma\otimes e+e\otimes\Gamma+
    \gamma_1\otimes\Gamma/\gamma_1+ \gamma_2\otimes\Gamma/\gamma_2
    +\gamma_3\otimes\gamma_4+ \gamma_4\otimes\gamma_3.
\end{array}
\label{eq:coproductEX1}
\eeq

Next, we calculate the antipode.  The lattices $\gamma_1,\gamma_2$ 
and $\gamma_4$ are primitive.  For $\gamma_3$, we have
\beq
\begin{array}{rl}
S(\gamma_3)&=-\gamma_3-S(\gamma_1)\gamma_3/\gamma_1
            -S(\gamma_2)\gamma_3/\gamma_2\\
	    &=-\gamma_3+2\gamma_1\gamma_2.
\end{array}
\eeq
We plug this in
\beq
S(\Gamma)=-\Gamma-\sum_{i=1,\ldots,4}S(\gamma_i)\Gamma/\gamma_i
\eeq
and get
\beq
\begin{array}{rl}
S(\Gamma)=&-\Gamma+\gamma_1\Gamma/\gamma_1 +\gamma_2\Gamma/\gamma_2 
+\gamma_3\Gamma/\gamma_3 \\
&-2\gamma_1\gamma_2\Gamma/\gamma_3+\gamma_4\Gamma/\gamma_4.\\
=&-\Gamma+\gamma_1\Gamma/\gamma_1 +\gamma_2\Gamma/\gamma_2 
-2\gamma_1\gamma_2\gamma_4+2\gamma_3\gamma_4.
\end{array}
\label{eq:antipodeEX1}
\eeq
namely, the lattice
\beq
\mbox{\epsfig{file=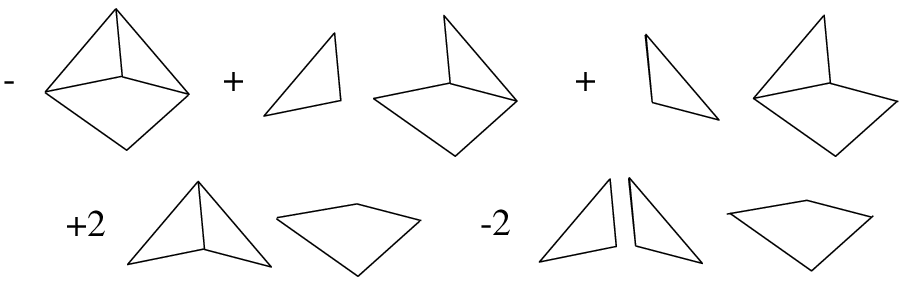}}
\!\!\!
\eeq
We can use (\ref{eq:coproductEX1}) and (\ref{eq:antipodeEX1}) to 
check that $O(\Gamma)=0$.

We can carry out the same calculations on the parenthesized weight 
$w_\Gamma$ in (\ref{eq:wEX1}) for this lattice.
The result is exactly the same as (\ref{eq:coproductEX1}) and 
(\ref{eq:antipodeEX1}), but we will write it out to indicate how 
parenthesized weights should be manipulated.  For the coproduct, we 
have
\beq
\begin{array}{rl}
\Delta(w_\Gamma)=&w_\Gamma\otimes 1 + 1\otimes w_\Gamma
+(w_{\gamma_1})\otimes\left((w_{\gamma_2})(w_{\gamma_4})\right)
+(w_{\gamma_2})\otimes\left((w_{\gamma_1})(w_{\gamma_4})\right)\\
&+\left((w_{\gamma_1})(w_{\gamma_2})\right)\otimes(w_{\gamma_4})
+(w_{\gamma_4})\otimes\left((w_{\gamma_1})(w_{\gamma_2})\right).
\end{array}
\eeq
The antipode is
\beq
\begin{array}{rl}
S(w_\Gamma)=&-w_\Gamma+\left(w_{\gamma_1}\right)\left(\left( 
w_{\gamma_2}\right)\left(w_{\gamma_4}\right)\right)
+\left(w_{\gamma_2}\right)\left(\left(w_{\gamma_1}\right) 
\left(w_{\gamma_4}\right)\right)\\
&+2\left(\left(w_{\gamma_1}\right)\left(w_{\gamma_2}\right)\right)
\left(w_{\gamma_4}\right) 
-2\left(w_{\gamma_1}\right)\left(w_{\gamma_2}\right) 
\left(w_{\gamma_4}\right).
\end{array}
\eeq


\subsection{The shrinking ``antipode''}

We can perform a renormalization group operation on a lattice by 
summing over possible values on some, or all, edges internal in the lattice, 
and so shrinking the lattice down to one with a smaller number of 
plaquettes, carrying effective couplings.  The partition function on 
the new lattice is a different function, on a different set of 
labels, than the original one.  When new and old couplings obey 
the renormalization group equation, the value of the effective partition 
function is equal to the  value of the original one. 

We will reproduce this by using the operations of the Hopf algebra we 
defined and get the correct effective couplings by using a modified 
version of the antipode $S$ of eq.\ \ref{eq:S}.  We will first give a 
general form of this operation, and then apply it to our $Z_2$ 
example.

First, we define an operation $R$ which: 1) when 
applied to a lattice $\Gamma$, it produces an effective lattice 
$R(\Gamma)$, {\em with the same external edges and their labels} as 
$\Gamma$.  $R(\Gamma)$ may also be a sum of lattices with the same 
external edges and labels as the original one.
2) when applied to a weight $w_\Gamma$ on that lattice it 
produces a new weight $R(w_\Gamma)$ on $R(\Gamma)$.  

We want $R$ to be a renormalization operation, which means that we 
want an equivalence relation
\bea
R(\Gamma)&\sim&\Gamma,\\
\label{eq:sim}
R(w_\Gamma)&\sim&w_\Gamma,
\label{eq:wsim}
\eea
which means that the two lattices are equivalent under renormalization.

Exactly what the relationship between the two weights is depends on 
the chosen coarse-graining scheme.  It is straightforward to state if 
$R$ is an {\em exact} scheme.  Then the partition 
function $Z_R(\Gamma)$ with weight $R(w_\Gamma)$ evaluates to the same 
number as the partition function $Z(\Gamma)$ with weight $w_\Gamma$, 
for all $w_\Gamma$ in the theory. 
This will be the case, for example, if we define $R$ to erase all 
internal edges on the lattice,
\beq
R(\Gamma)=\partial\Gamma,
\eeq
by summing over all labels on the erased edges:
\beq
R(w_\Gamma)=\sum_{\{q_i\in\bar{\Gamma}\}}w_\Gamma.
\eeq
($\bar{\Gamma}$ is the interior of $\Gamma$, as before). 
We will do a 
calculation of this kind of $R$ on a $Z_2$ lattice in the 
following subsection.  

In an {\em approximate} renormalization scheme, we expect that 
$R(w_\Gamma)\sim w_\Gamma$ if $Z(\Gamma)=Z_R(\Gamma)+Z_R^c(\Gamma)$, 
where $Z_R^c(\Gamma)$ is the correction terms, which should be 
appropriately small. 
 For example, $R(w_{\Gamma})$ may be truncation, or 
extraction of a pole term from $w_\Gamma$.  In a decimation scheme, 
$R$ will be an operation that chooses certain edges of $\Gamma$ to be 
the edges of the new lattice, with the original edge 
labels, and throws away the rest.  

For approximate $R$, we should note the following:  
$R(w_{\Gamma_1})\sim w_{\Gamma_1}$ does {\em not} imply 
$R(w_{\Gamma_1}) w_{\Gamma_2} \sim w_{\Gamma_1}w_{\Gamma_2}$. However, 
motivated by  \cite{Kreimer,KHA}, we will require that $R$ on two 
lattices $\Gamma_1=\prod_i\Gamma_1^i$ and $\Gamma_2=\prod_j\Gamma_2^j$ 
satisfies
\beq
R\left(\prod_i R(\Gamma_1^i) \prod_j \Gamma^j_2\right)=
\prod_i R(\Gamma_1^i)\prod_j R(\Gamma^j_2).
\eeq

Now define a shrinking operation $S_R$ as
\beq
S_R(\Gamma)=-R(\Gamma)-R\left(\sum_\gamma 
S_R(\gamma)\Gamma/\gamma\right)
\label{eq:SR}
\eeq   
on lattices, and
\beq
S_R(w_\Gamma)=-R(w_\Gamma)-R\left(\sum_\gamma 
S_R(w_\gamma)w_{\Gamma/\gamma}\right)
\label{eq:wSR}
\eeq 
on weights. 
This is a modification of the antipode (\ref{eq:S}).
The sum ranges over all proper sublattices in the given partition of 
$\Gamma$, as before, and stops when a primitive lattice is reached:
\beq
S_R(\gamma_p)=-\gamma_p.
\eeq
We will call $S_R$ the {\em shrinking antipode} (and sometimes refer 
to $S$ as the ``straight'' antipode).  

The equivalent of $O$ can be written down for $S_R$, both for lattices 
and for weights
\beq
O_R=m(S_R\otimes\id)\Delta.
\label{eq:OR}
\eeq
If $O_R$ evaluates to the same right hand side as $O$, namely $0$ for 
all $\Gamma\neq e$ and $0$ for all $w_\Gamma\neq 1$, then 
we will have a Hopf algebra for the renormalization group. 
By this we mean that $O_R$ is {\em equivalent} 
to 0 under the chosen $R$. 

For the weights in the following two examples, it is the case that
$O_R$ is equivalent to 0 under the chosen renormalization map $R$.  
First, we will give the
example of an exact block transformation on 
a $Z_2$ lattice, followed by an alternative definition for 
$S_R$, which applies to some coarse-graining schemes and is simpler. 
Next, in section 3, we give a simpler example than $Z_2$, the 
1-dimensional Ising model.

\subsection{Shrinking example}

In this example, we block transform the lattice (\ref{eq:labeled}) with 
partition (\ref{eq:example1}).

We will use the $R$ 
that eliminates all edges in the interior of the lattice:
\beq
R(\gamma)=\partial\gamma.
\label{eq:REX1}
\eeq
by summing over the labels on all the 
internal edges of $\gamma$:
\beq
R(w_\gamma)=
\sum_{q_i\in\bar{\gamma}=\pm 1} w_\gamma.
\label{eq:Rw}
\eeq
The sublattices
$\gamma_1, \gamma_2, \gamma_4$ are primitive.   For $\gamma_3$, we 
calculate (\ref{eq:SR})
\beq
S_R(\gamma_3)=-\partial\gamma_3+2\gamma_1\gamma_2, 
\eeq
which we plug in (\ref{eq:SR}) for $\Gamma$, and find
\beq
S_R(\Gamma)=-\partial\Gamma+\gamma_1\partial(\Gamma/\gamma_1)
+\gamma_2\partial(\Gamma/\gamma_2)+2\partial(\gamma_3)\gamma_4 
-2\gamma_1\gamma_2\gamma_4.
\label{eq:SREX1}
\eeq
namely, the lattice
\beq
\mbox{\epsfig{file=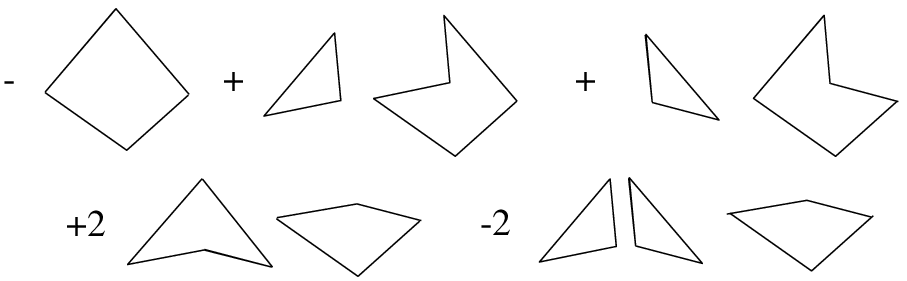}}
\eeq

The corresponding expression for weights needs:
\bea
&&
\begin{array}{rl}
&w'_{\gamma_3}:=R(w_{\gamma_3})
=e^{\kappa'_3 q_1 q_3 q_5 q_4},\\
&\mbox{with }\kappa'_3=
\tanh^{-1}\left(\tanh\kappa_1\tanh\kappa_2\right),
\end{array}\\
&&
\begin{array}{rl}
&w'_{\Gamma/\gamma_1}:=R(w_{\Gamma/\gamma_1})
=e^{\kappa'_2 q_2 q_3 q_7 q_6 q_4},\\
&\mbox{with }\kappa'_2=
\tanh^{-1}\left(\tanh\kappa_2\tanh\kappa_3\right),
\end{array}\\
&&
\begin{array}{rl}
&w'_{\Gamma/\gamma_2}:=R(w_{\Gamma/\gamma_2})
=e^{\kappa'_1 q_1 q_2 q_5 q_7 q_6},\\
&\mbox{with }\kappa'_1=
\tanh^{-1}\left(\tanh\kappa_1\tanh\kappa_3\right),
\end{array}\\
&&
\begin{array}{rl}
&w'_{\Gamma}:=R(w_{\Gamma})
=e^{\kappa' q_1 q_3 q_7 q_6},\\
&\mbox{with }\kappa'=
\tanh^{-1}\left(\tanh\kappa_1\tanh\kappa_2\tanh\kappa_3\right),
\end{array}
\eea
(we have used the 
method described in the footnote in Section 3)
and results in 
\beq
S_R(\Gamma)=-w'_\Gamma+(w_{\gamma_1})(w'_{\Gamma/\gamma_1})
+(w_{\gamma_2})(w'_{\Gamma/\gamma_2})+2(w'_{\gamma_3}) 
(w_{\gamma_4}) -2(w_{\gamma_1})(w_{\gamma_2})(w_{\gamma_4}).
\eeq

Substituting (\ref{eq:SREX1}) and ({\ref{eq:coproductEX1}) in 
(\ref{eq:OR}), we find
\beq
O_R(\Gamma)=\Gamma-\partial\Gamma+\gamma_1\partial(\Gamma/\gamma_1) 
-\gamma_1\Gamma/\gamma_1+\gamma_2\partial(\Gamma/\gamma_2)
-\gamma_2\Gamma/gamma_2
-\gamma_3\gamma_4+\partial(\gamma_3)\gamma_4,
\eeq
namely the lattice
\beq
\mbox{\epsfig{file=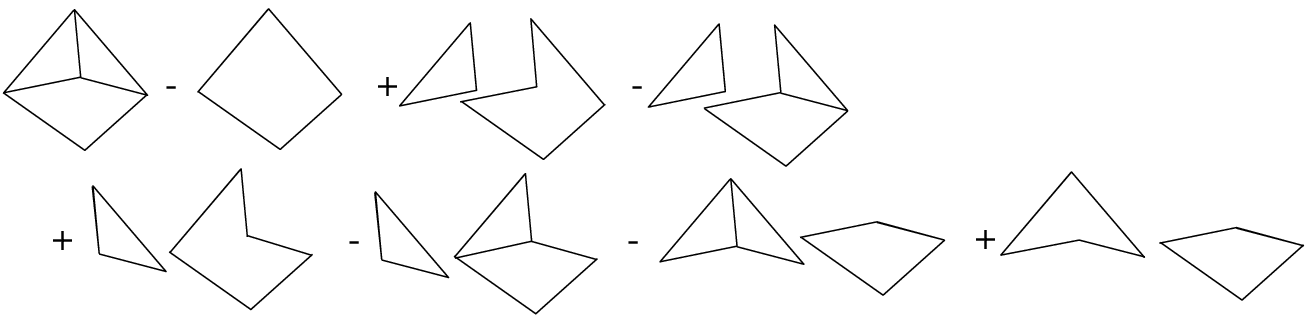}}
\eeq
This is  zero under the equivalence relation  
$\gamma\sim R(\gamma)=\partial\gamma$ in 
(\ref{eq:sim}) and (\ref{eq:REX1}).   
The corresponding antipode for weights also gives 
$O_R(w_\Gamma)=0$ under $w_\gamma\sim R(w_\gamma)$ and (\ref{eq:Rw}).

\subsection{An alternative definition of $S_R$ on the $Z_2$ lattice}
Keeping $R$ as above, we will now give a different definition of 
$S_R$, that can be used in an exact renormalization scheme like 
(\ref{eq:REX1}), (\ref{eq:Rw}).   
We will define 
\beq
S_R'(\Gamma)=-R(\Gamma)-\sum_\gamma S_R'(\gamma)\Gamma/\gamma,
\eeq
with the same expression also applying to the corresponding weights. 
It is different than $S_R$ since it is missing the overall $R$ 
operation.  Since, for our renormalization scheme (\ref{eq:REX1}), 
(\ref{eq:Rw}), 
$R(\Gamma_1\cdot\Gamma_2)=R(\Gamma_1)R(\Gamma_2)$, $S_R'$ differs from 
$S_R$ in the contributions of the remainders, which are not shrunk.

On our example, this gives
\beq
\begin{array}{rl}
    S'_R(\Gamma)=&-\partial\Gamma+\gamma_1\Gamma/\gamma_1 
    +\gamma_2\Gamma/\gamma_2 +\partial(\gamma_3)\Gamma/\gamma_3 \\
    &-2\gamma_1\gamma_2\gamma_4 +\gamma_4\Gamma/\gamma_4,
\end{array}
\label{eq:SRprime}
\eeq
which is the lattice
\beq
\mbox{\epsfig{file=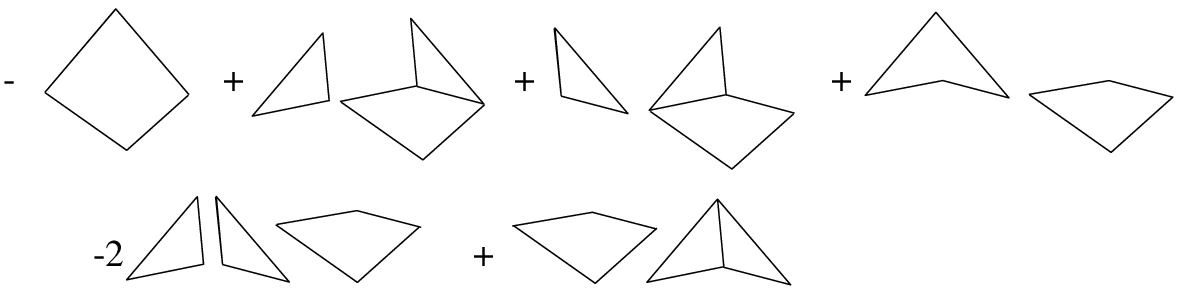}}
\eeq

Plugging (\ref{eq:coproductEX1}) and (\ref{eq:SRprime}) in $O_R$, we 
get
\beq
O_R(\Gamma)=\Gamma-R(\Gamma).
\eeq
Again this is zero under $\sim$.  

This definition of the shrinking antipode is attractive if we wish to 
use the second term in (\ref{eq:SRprime}) as an iterative equation, 
which calculates the effective weights on the shrunk lattice in terms 
of the weights on its sublattices.

\section{The Hopf algebra renormalization of the 1-d Ising model}

We now illustrate the method on a simple 1-dimensional example. 
Consider a finite Ising chain $\Gamma$.  This is a line, or a circle, 
with $N$ spins attached to it.  The coupling between spins 
$s_i$ and $s_j$ is $\kappa_{ij}$ when $i,j$ are adjacent sites and zero 
otherwise.   A Boltzmann weight for $\Gamma$ is 
\beq
w_{\Gamma}= \exp\left(\sum_{\langle i,j\rangle} \kappa_{ij} s_i 
s_j\right), 
\eeq
where $\langle i,j\rangle$ means that $i$ and $j$ are adjacent 
sites in the chain.  As before, we absorb $\beta$ in the other 
parameters.  Thus, the partition function for the system is 
\beq
Z(\Gamma)=\sum_{\{s_i\}} \exp(-\sum_{\langle i,j\rangle} \kappa_{ij} s_i 
s_j).  
\eeq

Clearly, partitioned Ising chains give rise to a Hopf algebra, as they 
are a special case of the 2-dimensional lattices we already analyzed.  
So do their Boltzmann weights. 
Given an allowed partition of $\Gamma$, a subchain 
$\gamma$  will either be a sequence of $n$ spins, 
with weight $w_{\gamma}=\prod_{i=k,\ldots,k+n} e^{\kappa_{ii+1} s_i 
s_{i+1}}$, or a disjoint union of such subchains, 
$\gamma_1\cup\gamma_2$, with weight $w_{\gamma_1} w_{\gamma_2}$.  
The remainder $\Gamma/\gamma$ is the 
subchain of $\Gamma$ that contains all spins except those internal 
in $\gamma$ and has weight $w_{\Gamma/\gamma}=w_{\Gamma}/w_{\gamma}$.  

The coproduct (\ref{eq:coproduct}), 
again produces all possible pairs of subchains and remainders in the 
given partition of $\Gamma$. We can easily identify the primitive 
elements.   There is only one type of primitive chain: 
a pair of adjacent spins.  A primitive weight then has the form  
$\gamma_p=e^{\kappa_{ij} s_i s_j}$.

To block transform $\Gamma$ using the Hopf algebra, we
first define the $R$ operation on such subchains as
\bea
R(\gamma)&=&\partial\gamma,\\
R(w_{\gamma})&=&
\sum_{\mbox{\small Internal spins of }\gamma=\pm1} w_\gamma.
         \label{eq:RIsing}
\eea
For  $\gamma=\gamma_1\cup\gamma_2$, 
$R(\gamma)=R(\gamma_1)R(\gamma_2)$ and, on a primitive diagram, 
$R(\gamma_p)=\gamma_p$.  

The details of the shrinking operation are best illustrated with examples.  The 
point can be made with an Ising model with 3 spins.  For larger 
chains, nothing new happens, except that the combinatorics produce 
more terms at each step. 

Consider then the chain 
\beq
	\begin{array}{c}\mbox{\epsfig{file=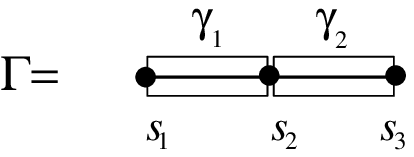}}\end{array},
\eeq
with weight $w_\Gamma=e^{\kappa_1 s_1 s_2}e^{\kappa_2 s_2 s_3}$, 
partitioned into subchains $\gamma_1, \gamma_2$, 
with $w_{\gamma_1}=e^{\kappa_1 s_1 s_2}$ and $w_{\gamma_2}=e^{\kappa_2 s_2 
s_3}$.  They are both primitive.  

The coproduct on $\Gamma$ is
\beq
\Delta[\Gamma]=
	   \Gamma\otimes e+ e\otimes\Gamma+ 
	       \gamma_1\otimes\gamma_2 +\gamma_2\otimes\gamma_1.
\eeq
The ``straight'' antipode $S$ of eq.\ (\ref{eq:S}), gives 
\beq
S(\Gamma)=-\Gamma+2\gamma_1\gamma_2.
\eeq
One can check that $O(\Gamma)=0$. 

The shrinking antipode $S'_R$ (eq.\ (\ref{eq:SRprime})) on $w_\Gamma$, gives
\beq
\begin{array}{rl}
S'_R\left(w_\Gamma\right)=& -R\left(w_\Gamma\right) + 
R\left(w_{\gamma_1}\right)w_{\gamma_2} +R\left(w_{\gamma_2}\right)w_{\gamma_1}\\
=& w_{\Gamma'}+2w_{\gamma_1}w_{\gamma_2},
\end{array}
\eeq
where $w_{\Gamma'}=R(w_\Gamma)$, which, using eq.(\ref{eq:RIsing}) 
is\footnote{
There are several ways to derive this formula and I will 
quickly outline one:\\
In $w(\Gamma)$, expand each factor using 
\beq 
e^{\kappa s_{i}s_{j}}= \cosh\kappa\ (1+x s_{i}s_{j}),
\eeq
with $x$ given by $x=\tanh(\kappa)$.
Then, keep $s_{1}$ and $s_{3}$ fixed and sum over 
$s_2=\pm 1$.  Only terms with even powers of internal spins 
survive, and we are left with 
\beq
2\cosh\kappa_1\cosh\kappa_2\left(1+(x_1 x_2)s_{1}s_{3}\right).
\eeq
We ignore the factor $2$ as it does not affect the calculation of 
any expectation values, and note that this is a nearest-neighbour 
interaction $e^{\kappa' s_1 s_{3}}$ if
we redefine the coupling to be $\kappa'$ given by equation 
(\ref{eq:kappa'}).  } 
\beq
w_{\Gamma'}=e^{\kappa'}s_1 s_3, 
\qquad
\mbox{with }\kappa'=\th^{-1}\left(\tanh\kappa_1\tanh\kappa_2\right).
\label{eq:kappa'}
\eeq

Therefore,  $O_R(w_\Gamma)=-w_{\Gamma'}+w_{\Gamma}$.  That is, the 
fully blocked chain ${\Gamma'}$ is:
\beq
w_{\Gamma'}=w_\Gamma-O_R(w_{\Gamma}).
\label{eq:OIsing}
\eeq

For a larger Ising chain,  
either $w_{\Gamma'}=e^{K s_1 s_N}$, if $\Gamma$ is an open chain 
with external spins $s_1$ and $s_N$ (and $K$ is the overall effective 
coupling, the generalization of (\ref{eq:kappa'}) to $N$ spins), 
or $w_{\Gamma'}=e^{K s_i^2}$ 
for some spin $s_i\in\Gamma$, if $\Gamma$ is a closed chain.  This can 
be thought of as the ``fully block transformed'' $\Gamma$, or the 
evaluation of $w(\Gamma)$.

An Ising/Potts model in two dimensions works in the same way. 

It is important to note the following.  In this model, as well as in 
the $Z_2$ case, 
eq.(\ref{eq:OIsing}) is redundant in the 
calculation of the fully block transformed chain ${\Gamma'}$,
since we have 
already calculated it as the term $R(w_\Gamma)$ in $S_R(w_\Gamma)$.  
However, this is a special property of these models and the exact 
renormalization scheme $R$ that we employed.  It will not be the case, 
for example, in spin foam models, where we do need to calculate $O_R$.  
We discuss this in the next section.


\section{Basics of the Hopf algebra renormalization of a 1+1 spin foam} 

We now show that a similar Hopf algebra is defined on 1+1 spin foams 
and can be used in the renormalization of spin foam models.  In 1+1 
dimensions, a spin foam $\Gamma$ is a 2-dimensional lattice, with 
vertices $v$ and faces $f$.  The faces are labeled by unitary 
irreducible representations $a_f$ of a Lie group $G$.  $\dim a_f$ is 
the dimension of the representation. Each vertex $v$ is labeled by an 
amplitude $A(v)$, a function of the labels on the faces
adjacent to that vertex.  Particular choices of the group and these 
functions give rise to specific spin foam models \cite{euclideanSF}.
Also, the faces of $\Gamma$ may be labeled by integers representing 
geometric properties such as lengths or matter\cite{ALA}.

The partition function for a spin foam has the form 
\beq
Z=\sum_\Gamma N(\Gamma)\sum_{
\begin{array}{c}
\mbox{\footnotesize Labelings}\\
{\mbox{\footnotesize{on }} \Gamma}
\end{array}
}
\prod_{f\in\Gamma}\dim a_f\prod_{v\in\Gamma}A_v.
\label{eq:Zsf}
\eeq
where the first sum ranges over all spin foams that extrapolate 
between fixed initial and final spin networks. 
We will treat the second sum in $Z$ as a generalization of 
a lattice gauge theory and list the basic features of its 
renormalization by the Hopf algebra method. 

We will call ``subfoam'' a proper sublattice $\gamma$ of a spin foam 
$\Gamma$ (one that is not empty and not $\Gamma$ itself).  As in the 
case of the $Z_2$ lattice gauge theory, we will work with {\em 
partitioned} spin foams, namely spin foams which have been marked by a 
partition into subfoams in which no two subfoams overlap.

Let $A$ be the collection of partitioned 1+1 spin foams.  Since each spin 
foam can be multiplied by a complex number, we will think of $A$ as an 
algebra over the complexes.  {\em Multiplication} is the disjoint 
union of two spin foams: $\Gamma_1\cdot\Gamma_2=\Gamma_1\cup\Gamma_2$.  
Denoting the empty spin foam by $e$, $\Gamma\cdot e=e\cdot \Gamma$, 
for every spin foam $\Gamma$.  The {\em unit} operation is the map 
$\epsilon:{\bf C}\rightarrow A$, which for some complex number $c$ 
gives $\epsilon(c)=ce$.   At least in 1+1 dimensions, $A$ is a 
commutative algebra since the order of two disjoint spin foams does 
not matter.

Two further operations that are natural for spin foams turn $A$ into a 
coalgebra.  First, the {\em counit} annihilates all spin foams, except 
the empty one:
\beq
\coe(\Gamma)=
\left\{ \begin{array}{lr}
0& \mbox{for }\Gamma\neq e,\\
1& \mbox{for }\Gamma=e.
\end{array}\right.
\eeq

The remainder $\Gamma/\gamma$ 
of a subfoam $\gamma$ is the subfoam obtained by shrinking $\gamma$ 
to a point in $\Gamma$:
\beq
	\begin{array}{c}\mbox{\epsfig{file=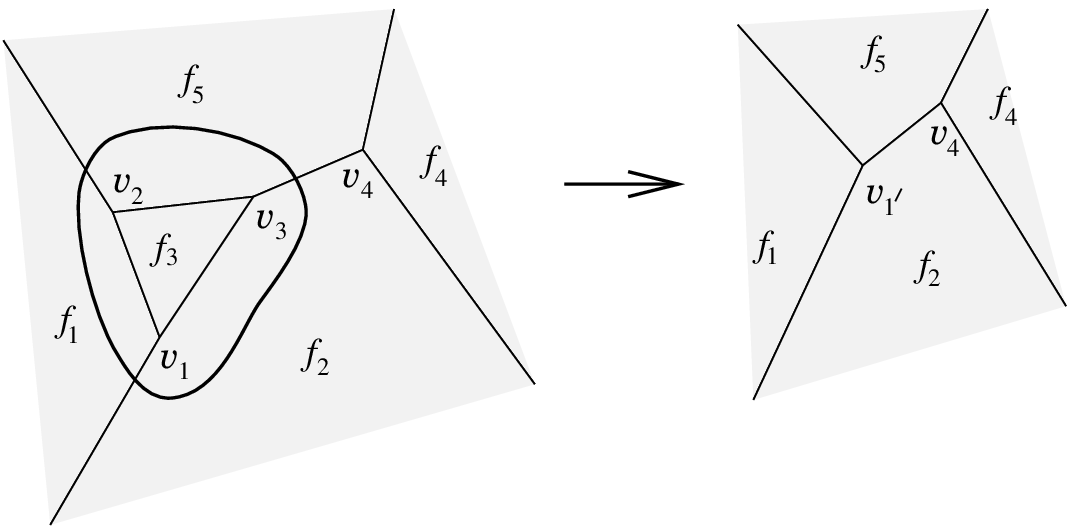}}\end{array}.
\label{eq:spinfoam}
\eeq
Note that this is different than the 
remainder we used in section 2, where we ``cut out'' $\gamma$.  
Rather, it is the same as Kreimer's remainder in \cite{Kreimer}.

The {\em coproduct} splits a spin foam into a sum of all its possible 
subfoams, paired to their remainders:  
\beq
\Delta(\Gamma)=\Gamma\otimes e+e\otimes\Gamma+ \sum_\gamma 
\gamma\otimes\Gamma/\gamma.
\eeq
As before, $\gamma$ in the above sum ranges over all subfoams in the 
given partition.  
For a spin foam $\Gamma$ with no subfoams, i.e.\ a {\em primitive spin 
foam}, we have
\beq
\Delta(\Gamma)=\Gamma\otimes e+e\otimes\Gamma.  
\eeq
If there is a restriction in the valence of the spin foam vertices in 
a given spin foam model, 
then there is a finite set of primitive spin foams\footnote{
These primitive foams can be compared to Feynman diagrams with no 
subdivergences.  For a given theory, for example $\phi^4$, 
we can list the diagrams with no subdivergences, of 
which there is a finite number. }.

The straight antipode on spin foam diagrams is  $S$ as given in 
eq.(\ref{eq:S}), but with the remainder defined above.  

All of the above operations on the spin foam complexes have their 
counterparts for the spin foam weights, as in the 
lattice gauge theory we have already studied in detail. Thus, 
in the above example (\ref{eq:spinfoam}), the weight for $\Gamma$ is 
\beq
w_\Gamma=\prod_{i=1,\ldots,5}\dim a_{f_i} \prod_{k=1,\ldots,4} A_{v_k},
\eeq
while the marked subfoam has weight
\beq
w_\gamma=\prod_{i=1,2,3,5}\dim a_{f_i}\prod_{k=1,2,3}A_k.
\eeq

For the shrinking antipode $S_R$, we need to define a renormalization 
recipe $R$ that provides effective vertices in terms of the original 
ones.  We have no explicit renormalization scheme to suggest in this 
paper, so we will simply note three basic things.  One is that a 
possible $R$ operation is the recoupling moves.  For example, it is 
possible to shrink the subfoam in the above example using a 3-to-1 
move.  

Second, if the theory is triangulation invariant the 
effective vertices contain no information about the ones that we have 
shrunk.  This makes the algebra trivial. 
 
Finally, we note that the equivalence relation 
$R(\Gamma)\sim\Gamma$ modifies the summation over all interpolating 
spin foams in the partition function.  In general, there is an 
infinite number of such foams, which makes it difficult to handle this 
sum in any context other than triangulation invariance or renormalized 
spin foams.  However, we can split the first sum in (\ref{eq:Zsf}) 
into sums over spin foams that are equivalent under renormalization.
Developing a particular scheme to understand the renormalization group 
flow would then enable us to calculate the partition function. 

\section{Discussion}

We have expressed block spin transformations of spin systems and spin 
foam as an equivalence relation and a modification of the antipode of 
a Hopf algebra, originally used by Kreimer for the perturbative 
renormalization of quantum field theory.

As a method to carry out the renormalization group transformation of 
spin systems, this is promising especially on inhomogeneous lattices, 
as it can efficiently keep track of the combinatorial part of the 
problem.  The antipode of the algebra, which produces the generalized 
renormalization group equation, is an iterative equation and thus 
ready to be implemented numerically.  (Broadhurst and Kreimer easily  
calculated 4d Yukawa theory to 30 loops using the perturbative form of this 
algebra \cite{BK}.)  

However, the summations involved in a single block 
transformation may still be formidable.  
In sections 2 and 3, we discussed exact renormalization schemes of 
spin systems.  In fact, the power of the algebra is with approximate 
schemes, such as truncations of the Boltzmann weights.  This is also 
the case for spin foams.  Such schemes will be studied in future 
work.  

We have given the operations of the algebra on the generic spin foam 
partition function.   The equivalence relation 
defined via this algebra defines equivalence classes of spin foams,
corresponding to the same effective vertex, and  we argued that this 
may be used to reduce the number of spin foams to be summed over 
in the partition function. 

As we discussed in 2.3, if the condition $O_R$ is satisfied for every
weight in the algebra, then the shrinking antipode is a genuine
antipode.  This would imply that the  renormalization
group equation can be embedded in this Hopf algebra. 
For the choices of weights in the examples we studied, this is
satisfied.   Whether this is generally the case
for spin systems or gauge systems is left for further work. 

We should note that our discussion applies to 
euclidean spin foams, as we have not paid attention to the orientation 
of the edges of the spin foam.  However, we expect that the 
construction can also be applied to the causal spin foams 
\cite{lorentzianSF}.
Also, although we have given the operations on a 1+1 spin foam, there 
should be no obstruction to an analogue in higher dimensions.  
The only necessary ingredient in this Hopf algebra is the rooted tree 
structure (the partitioned lattices), which of course exists in 
higher dimensions.  

To determine whether the method here is a useful tool in spin foam 
renormalization, it is of course necessary to apply it to specific 
spin foam models.  As we said above, topological state sum models 
appear unsuitable as the $R$ operation is trivial.  A good candidate 
is the Ambjorn-Loll-Anagnostopoulos Lorentzian gravity model in 2 
dimensions, as it has a transfer matrix formulation and its continuum 
limit is known.


\section*{Acknowledgments}

I am grateful to Eli Hawkins, Des Johnson, Shahn Majid and Lee Smolin for useful 
discussions, and the hospitality of Chris Isham and the Theory 
Group at Imperial College, where this work was carried out.  This work 
was supported by NSF grants PHY/9514240 and PHY/9423950
to the Pennsylvania State University and a gift from the Jesse 
Phillips Foundation.


\end{document}